\documentclass[rnote,structabstract]{aa}
\usepackage{graphicx}
\usepackage{txfonts}

\usepackage{xspace}
\usepackage{verbatim} 
\usepackage{natbib}
\bibpunct{(}{)}{;}{a}{}{,} 
\newcommand{\usun}{\ensuremath{_{\odot}}\xspace}
\newcommand{\eff}{\ensuremath{_{\rm eff}}\xspace}

\newcommand{\ong}{\ensuremath{\sim}\xspace}

\newcommand{\st}{\ensuremath{_*}\xspace}
\newcommand{\um}{\ensuremath{\mu\rm m}\xspace}

\newcommand{\grg}{\ensuremath{\rm f_{mm}}\xspace}

\newcommand{\e}[1]{\ensuremath{\rm ~\times~10^{#1}}\xspace}
\newcommand{\ten}[1]{\ensuremath{\rm 10^{#1}}\xspace}

\begin{document}
  \title{Full two-dimensional radiative transfer modelling of the transitional disk LkCa 15}
  \titlerunning{Full two-dimensional radiative transfer modelling of the transitional disk LkCa 15}
  \authorrunning{Mulders, Dominik, Min}
  \author{G.D.Mulders\inst{1,3} \and C. Dominik\inst{1,4} \and M. Min \inst{2}}
  \institute{
    Astronomical Institute ``Anton Pannekoek'', University of Amsterdam, PO Box 94249, 1090 GE Amsterdam, The Netherlands
    \and 
    Astronomical institute Utrecht, University of Utrecht, P.O. Box 80000, NL-3508 TA Utrecht, The Netherlands
    \and 
    SRON Netherlands Institute for Space Research, PO Box 800, 9700 AV, Groningen, The Netherlands
    \and 
    Afdeling Sterrenkunde, Radboud Universiteit Nijmegen, Postbus 9010, 6500 GL Nijmegen, The Netherlands
}
  \date{Received 23 June 2009 / Accepted 05 Jan 2010}
  \offprints{G.D.Mulders, \email{mulders@uva.nl}}

  \abstract
{With the legacy of Spitzer and current advances in (sub)mm astronomy, a considerable number of so-called 'transitional' disks has been identified which are believed to contain gaps or have developped large inner holes, some filled with dust. This may indicate that complex geometries may be a key feature in disk evolution that has to be understood and modelled correctly. 
The disk around \object{LkCa 15} is such a disk, with a large gap ranging from $\sim $5 - 46 AU, as identified by Espaillat et al. (2007) using 1+1D radiative transfer modelling. To fit the spectral energy distribution (SED), they propose two possible scenarios for the inner ($<$5 AU) disk - optically thick or optically thin - and one scenario for the outer disk.}
{We use the gapped disk of LkCa 15 as a case in point to illustrate the importance of 2D radiative transfer in transitional disks by showing how the vertical dust distribution in dust-filled inner holes determines not only the radial optical depth but also the outer disk geometry.}
{We use MCMax, a 2D radiative transfer code with a self-consistent vertical density and temperature structure, to model the SED of LkCa 15.}
{We identify two possible geometries for the inner \textit{and} outer disk that are both different from those in Espaillat et al. (2007). An inner disk in hydrostatic equilibrium reprocesses enough starlight to fit the near infrared flux, but also casts a shadow on the inner rim of the outer disk. This requires the outer disk scale height to be high enough to rise out of the shadow. An optically thin inner disk does not cast such a shadow, and the SED can be fitted with a smaller outer disk scale height. For the dust in the inner regions to become optically thin however, the scale height would have to be so much higher than its hydrostatic equilibrium value that it effectively becomes a dust shell. It is currently unclear if a physical mechanism exists which could provide for such a configuration.}
{We find that the radial optical depth of dust within the inner hole of LkCa 15 is controlled by its vertical distribution. If it turns optically thick, the outer disk scale height must be increased to raise the outer disk out of the inner disk's shadow.}
     
      \maketitle
      \section{Introduction}
      Protoplanetary disks are the main sites of planet formation. Within them, small dust grains grow from micron to millimeter sizes and settle to the midplane where they are believed to eventually form planetary systems. Sufficiently massive planets locally deplete the disk of gas and dust and may open a gap or enlarge the inner hole. Recently, considerable numbers of so-called transitional disks have been identified on the basis of their low near infrared excess \citep{2007MNRAS.378..369N}, indicating that their inner regions may be (partially) depleted of dust.

      \object{LkCa 15} is such a transitional disk, but with a more complex geometry: It has an inner disk, a gap and an outer disk. The outer disk, seen at an inclination of 51\degr, extends from 46 to 800 AU \citep{2006A&A...460L..43P}. The Spitzer data show that the inner hole is not devoid of dust between 0.1 and 5 AU \citep[][hereafter Esp07]{2007ApJ...670L.135E}. On the basis of 1+1D RT modelling, Esp07 identify two possible scenarios for the inner regions: one where the dust is optically thick and one where it is optically thin. Additional observations show that this dust is most likely present in the form of a small optically thick inner disk at the dust evaporating radius \citep[][ hereafter Esp08]{2008ApJ...682L.125E}. The gap inbetween is mostly empty.

      In this paper we will model both geometries proposed by Esp07 using a full 2D radiative transfer code and focus on the vertical structure of the inner \textit{and} outer disk. Our modelling approach differs from that of Esp07 in one important point: We combine the several disk components in one model \textit{before} we perform the radiative transfer and create the spectral energy distribution (SED), whereas they create the SED \textit{after} performing the radiative transfer on each seperate component. Although their approach is valid if the seperate components do not influence each other, we will show that at least for LkCa 15 this is not the case.

      \section{Disk model}
      The disk model we use here is MCMax \citep{2009A&A...497..155M}, a 2D Monte Carlo radiative transfer code with a selfconsistent vertical structure. The Monte Carlo radiative transfer simulation is based on that of \citet{2001ApJ...554..615B} and calculates the dust temperature for a given density structure. It also calculates the vertical structure by directly integrating the equation of hydrostatic equilibrium using the Monte Carlo temperature. Because the vertical structure calculation requires the temperature in the midplane to be accurately determined, MCMax treats those regions with an analytical diffusion approximation (see \cite{2009A&A...497..155M} for details). By iterating between the Monte Carlo simulation and the hydrostatic equilibrium calculation a self-consistent solution is found. This typically takes four or five iterations. MCMax also includes a modified random-walk approximation that speeds up the calculations in regions with extreme optical depths by making multiple interaction steps in a single computation (\citet{2009A&A...497..155M}, see also \citet{1984JCoPh..54..508F}). The typical run time is less than three minutes on a single core CPU for a model with a radial optical depth in the midplane of $\tau_{\rm 1\mu m}=10^6$ using six iterations with \ten{5} photon packages each.
      
      Because models with well-mixed dust and gas in vertical hydrostatic equilibrium tend to overpredict the disk's surface height and far infrared flux, dust settling and grain growth are necessary ingredients to fit SEDs \citep[e.g.][]{2006ApJ...638..314D}. For the present study we will describe them with two parameters that lower the disk surface: The mass fraction of large grains \grg and a sedimentation parameter $\Psi_{\rm all}$. The sedimentation parameter $\Psi_{\rm all}$ describes the settling of small dust particles and is implemented by \textit{reducing the dust scale height with respect to the selfconsistent gas scale height}: $\rm{H_{dust}(z)} = \Psi_{\rm all} \rm{H_{gas}(z)}$. A similar global sedimentation parameter has also been used by \citet{2007A&A...471..173R}. Both parameters have a different effect on the surface shape and SED and are not degenerate. The sedimentation parameter is global and reduces the scale height in both the inner and outer disk and therefore scales the SED accordingly. The fraction of large, mm-sized grains only affects the surface height in the outer disk: Increasing \grg reduces the flaring angle of the outer disk and mostly affects the far infrared flux.
      
      We will use a bimodial size distribution here, consisting of small (0.1\um) and big (2 mm) grains. For the dust opacity of the small particles we use that of 0.1 \um grains, consisting of 80\% silicate \citep{1995A&A...300..503D} and 20\% amorphous carbon \citep{1993A&A...279..577P}. We do not model them as compact spheres, but as irregular shaped particles \citep{2005A&A...432..909M}. The opacity of the big particles is modelled using 2 mm sized irregular shaped silicate particles. The total opacity is determined by the fitting parameter \grg, the fraction of mass in big particles. To settle the big grains to the midplane, we introduce a stratification parameter $\Psi_{\rm big}$ that reduces the scale height of large grains with respect to that of small grains: $\rm{H_{big}(z)} = \Psi_{\rm big} \rm{H_{dust}(z)}$.

      For the stellar photosphere we use a main-sequence Kurucz model with the stellar mass, luminosity and effective temperature from \citet{1996yCat..21010117K} (Table \ref{tab:model_parameters}). This photosphere is fitted to the de-reddened photometry between 0.5 \um and 2 \um by scaling the distance.
       
      \section{Results}
\begin{figure}
	\centering
	\framebox[\linewidth]{
	  \begin{minipage}{\linewidth}
	    \centering
	  \includegraphics[width=0.95\linewidth]{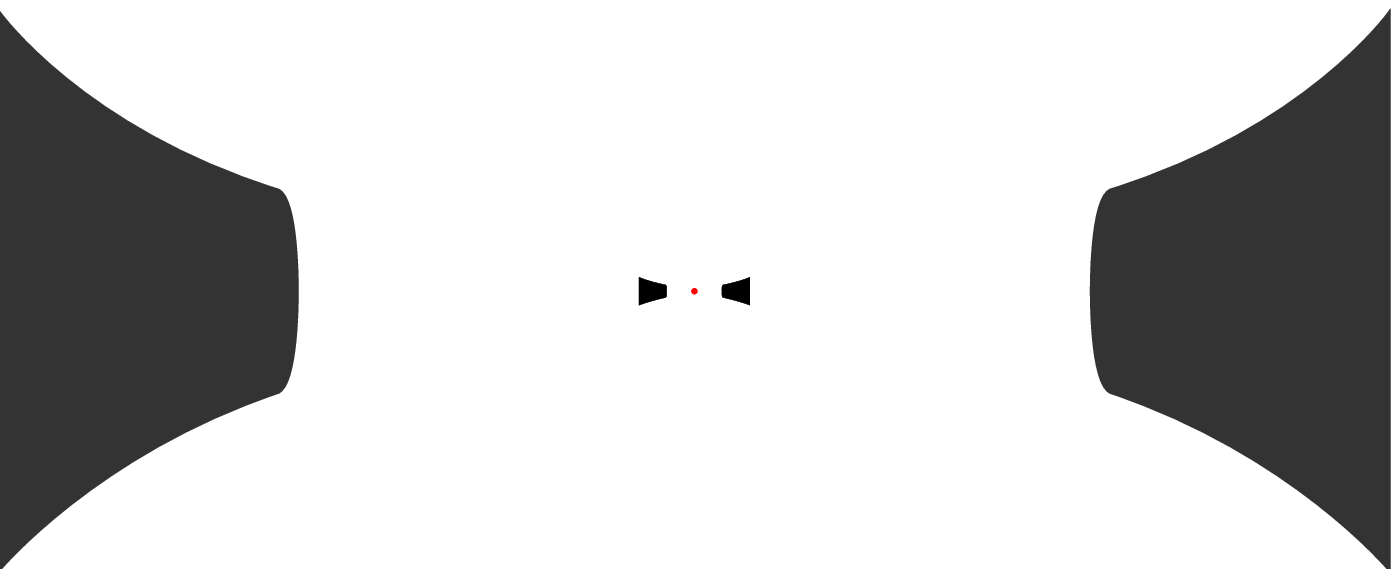}
	  \put(-170,90){a) Optically thick inner disk} \\
	  \vspace{0.5cm}
	  \includegraphics[width=0.95\linewidth]{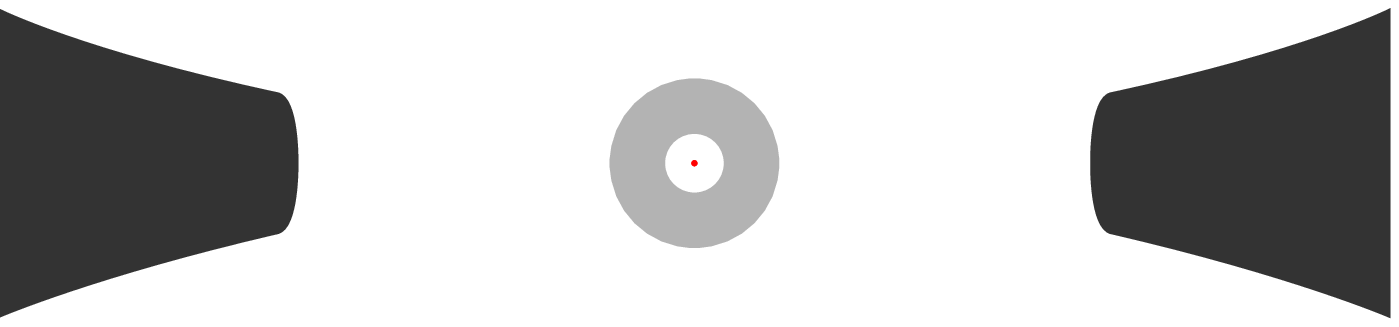}
	  \put(-170,90){b) Optically thin dust shell}
	  \end{minipage}
	}
	\framebox[\linewidth]{
	  \includegraphics[width=0.95\linewidth]{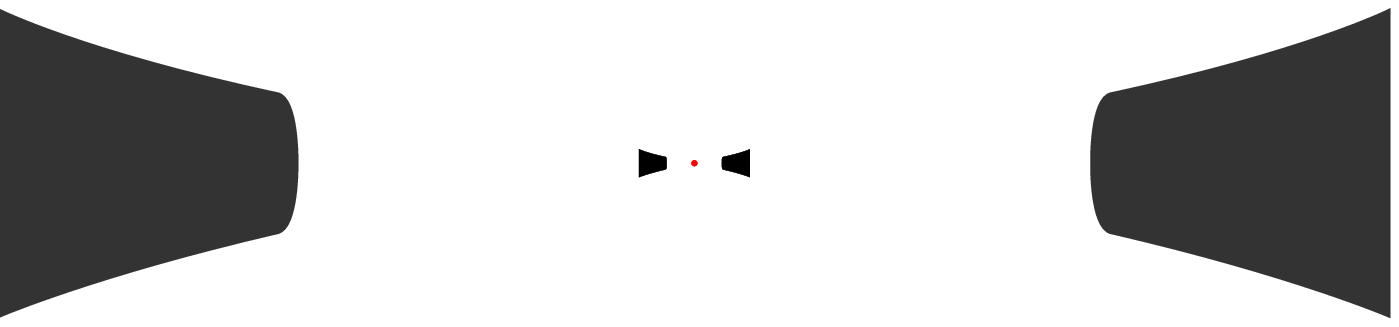}
	  \put(-135,90){c) Esp07}
	}
	\caption[]{Sketches (not to scale) of the two possible geometries for LkCa 15 from this paper (top panel) and - for comparison - that from Esp07 (bottom panel). 

Depicted are: a) A model with a 1 AU optically thick inner disk and an outer disk that rises out of its shadow. b) A model with a 5 AU optically thin spherical dust shell and a flatter outer disk with a fully illuminated inner rim. c) The model proposed by Esp07, a combination of the inner disk from (a) with the fully illuminated outer disk of (b).
	  \label{fig:LkCa15_sketch}
	}
      \end{figure}

      For each of the geometries (Fig. \ref{fig:LkCa15_sketch}) we follow a different approach. In the optically thick case, we model the inner disk first to constrain the size of the shadow cast on the outer disk and proceed by modelling the outer disk. Note that although we change the fit parameters for one component at a time, we always perform the radiative transfer including all components. In the optically thin case, the inner disk produces no shadow and we can model the outer disk first. We then proceed by including an inner disk so we can explore the effect of the vertical dust distibution on the radial optical depth.
     
      \subsection{Optically thick inner disk}

      \begin{figure}
	\includegraphics[width=\linewidth]{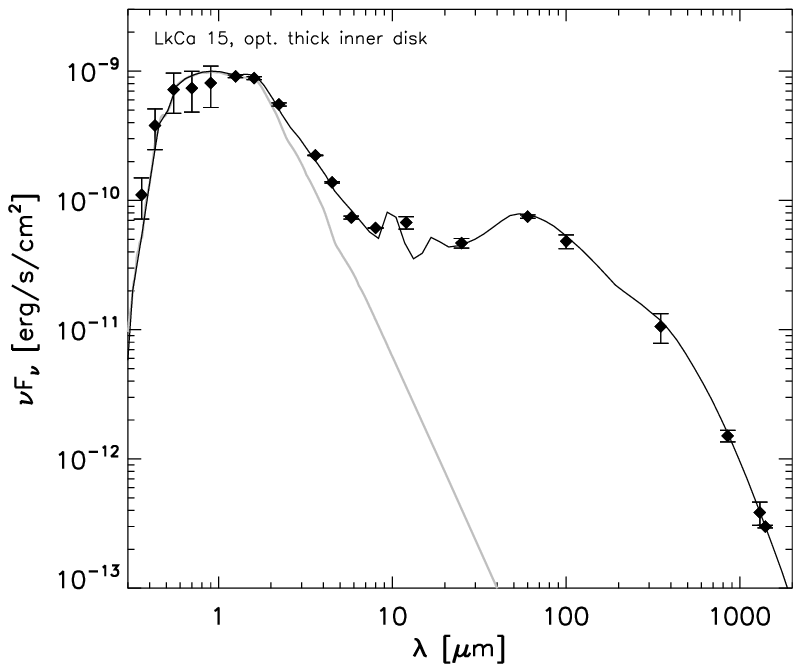} 
	\includegraphics[width=\linewidth]{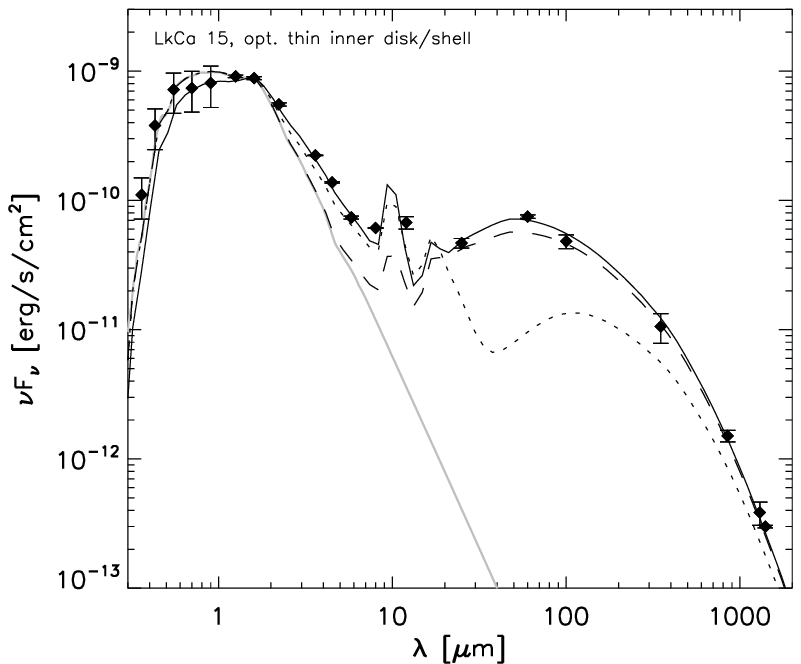} 
      \caption[]{Spectral Energy Distribution of LkCa 15 (diamonds) with 1$\sigma$ errorbars. Overplotted in grey is the stellar photosphere. The top panel shows our best-fit disk model with an optically thick inner disk (Fig. \ref{fig:LkCa15_sketch} a)

	The solid line in the bottom panel shows our best fit with an optically thin dust shell of 1\e{-11} M\usun (Fig. \ref{fig:LkCa15_sketch} b). The dotted line shows the same outer disk combined with an inner disk in hydrostatic equilibrium of 2\e{-10} M\usun that is optically thin in the \textit{vertical} direction (Fig \ref{fig:LkCa15_sketch} c). It reproduces the near and mid infrared flux, but is not radially optically thin. It casts a shadow on the outer disk and underpredicts the far infrared flux by almost a factor of ten. The dashed line shows the same model with a very flat (z/r \ong 0.01) inner disk. This minimizes the size of the shadow, but such an inner disk does not reprocess enough of the stellar radiation to fit the flux shortwards of 20 \um.
	References - 
	(UBVRI) \citet{1996yCat..21010117K}; 
	(JHK) \citet{2006AJ....131.1163S}; 
	(Spitzer IRAC) \citet{2007ApJS..169..328R}; 
	(IRAS) \citet{1992ApJS...78..239W}; 
	(submm) \citet{2005ApJ...631.1134A}; 
	(mm) \citet{2006A&A...460L..43P}; 

	  \label{fig:LkCa15_SED}}
      \end{figure}

      \begin{figure}
	\includegraphics[width=\linewidth]{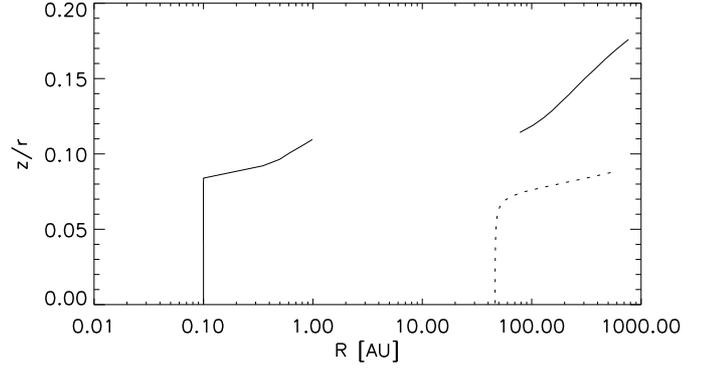}
	\caption[]{Radial $\tau$=1 surfaces for our best-fit disk models of LkCa 15. The solid line shows the disk model with an optically thick inner disk. The inner disk casts a shadow on the outer disk, and only the disk surface is directly illuminated. The dotted line shows the disk model with an optically thin dust shell. The inner rim of the outer disk is fully illuminated, and the outer disk is much flatter. In this plot, photons from the central star travel on horizontal lines (constant z/r).
	  \label{fig:LkCa15_tau}}
      \end{figure}
      In the optically thick case, the dust mass in the inner disk is hard to constrain from the SED. Therefore we use the same surface density powerlaw (p=1) as for the outer disk, which yields $\Sigma(1AU) = 560\ \rm{g/cm^2}$, assuming a dust-to-gas ratio of 1:100. We fix the inner radius at the dust evaporating radius at 0.1 AU and fit an outer radius for the inner disk of 1.0 AU, yielding a dust mass for the inner disk of 4\e{-6} M\usun. Our best fit (Fig. \ref{fig:LkCa15_SED}) does not require any sedimentation, and does not constrain stratification or big particle fraction, for which we will use the fit values from the outer disk.

      This choice of inner disk parameters is not unique, but the size of the shadow cast on the outer disk is. The height at the outer edge of the inner disk (z/r = \ong 0.1, Fig. \ref{fig:LkCa15_tau}) sets the total amount of energy reprocessed by the inner disk, which is constrained by the SED. The size of the shadow is therefore approximately equal for all solutions and independent of the exact inner disk geometry.

      With the size of the shadow cast by the inner disk constrained, we proceed by fitting the outer disk. We find a disk mass of 0.3 M\usun using p=1. This disk mass is much higher than previously reported, for instance by \citet{2005ApJ...631.1134A}. This can be attributed to the millimeter opacity of our big particles, which is a factor of five lower at 1.3 mm. Corrected for this, the disk mass for \object{LkCa 15} does not differ with from that of \citet{2005ApJ...631.1134A}.

      We lower the outer disk surface by using a sedimentation parameter of $\Psi_{\rm all}$=0.35 and a big particle fraction of 94\%. We also settle the big grains to the midplane ($\Psi_{\rm big}$=0.5) to fit the relatively low flux at 350 \um. A model without stratification ($\Psi_{\rm big}$=1) overpredicts this flux a little, but is still consistent with the large errors at that wavelength. The inner rim of the outer disk is almost completely shadowed by the inner disk, and only the disk surface is illuminated (Fig. \ref{fig:LkCa15_tau}). The reduction in surface height compared to a well-mixed model without big particles is four, comparable to that of other T-Tauri stars \citep{2006ApJ...638..314D}.

      \subsection{Optically thin inner disk}
      To investigate the effects of the optical depth of the inner disk on the outer disk structure, we first construct a disk model without an inner disk. In this model, the inner rim of the outer disk is fully illuminated, and its contribution to the SED from 25\um to 100\um will be much larger compared to a shadowed outer disk. To fit the far infrared flux, we need stronger sedimentation ($\Psi_{\rm all}$=0.25) and more big particles (\grg=99\%), so the outer disk becomes twice as flat (Fig. \ref{fig:LkCa15_tau}, dotted line). We find a total reduction in the surface height of a factor of eight, significantly higher than in other T-Tauri stars, which was also reported by Esp07.

      To fit the complete SED, we follow Esp07 and add an inner disk between 0.1 and 5 AU. If we assume this disk is in vertical hydrostatic equilibrium, we fit a dust mass in small grains of 2\e{-10} M\usun as constrained by the flux in the near and mid infrared. This inner disk however is not optically thin to starlight. Its radial $\tau$=1 surface reaches the same height (z/r$\sim$0.1) as the outer disk, casting a shadow over the entire outer disk which results in a far infrared flux drop by a factor of ten (dotted line in the lower panel of Fig. \ref{fig:LkCa15_SED}).

      One way to minimize the effect of this shadow is to make the inner disk extremely flat (z/r $\sim$ 0.01) (dashed line in the lower panel of Fig. \ref{fig:LkCa15_SED}). Although this model fits the far infrared flux reasonably well, the inner disk does not reprocess enough of the stellar radiation to reproduce the near infrared excess. Another way to avoid the shadow is to increase the disk scale height much beyond its hydrostatic equilibrium value. By spreading the same amount of dust over a larger solid angle, its radial optical depth decreases. For the material to become optically thin, the scale height has to be so high that we will model it as a spherical shell around the star (solid line in the lower panel of Fig. \ref{fig:LkCa15_SED}), but the exact geometry is hard to constrain as long as the material remains optically thin. We need only 1 \e{-11} M\usun of dust in this shell because it does not shield itself and is therefore hotter. It is also in our line of sight and reddens the starlight by an additional A$_v$=0.25. 
      %
      %
      \begin{table}
	\title{Model parameters}
	\centering
	\begin{tabular}{llllll}
	  \hline\hline
	  Parameter  &\multicolumn{2}{c}{Value} & Range \\
	  \hline
	  T\eff [K]      & \multicolumn{2}{c}{ 4350$^{1}$}  \\
	  L\st [L\usun]  & \multicolumn{2}{c}{ 0.74$^{1}$}  \\
	  M\st [M\usun]  & \multicolumn{2}{c}{ 0.97$^{2}$}  \\
	  \hline
	  d [pc]                   & \multicolumn{2}{c}{126}  & 110-140      \\ 
	  R$_{\rm in,outer}$ [AU]  & \multicolumn{2}{c}{46}  \\
	  R$_{\rm out,outer}$ [AU] & \multicolumn{2}{c}{800} \\
	  M$_{\rm disk,outer}$ [M\usun] & \multicolumn{2}{c}{0.3}  & 0.01-1.0 \\
	  p                        & \multicolumn{2}{c}{1.0}        \\
	  R$_{\rm in,inner}$ [AU]  & \multicolumn{2}{c}{0.1} \\
	  \hline
	                 & Optically thick     & Optically thin       \\
	                 & inner disk          & dust shell          \\
	  \hline	  
	  R$_{\rm out,inner}$ [AU] & 1              & 5           & 0.2-5 \\
	  M$_{\rm dust,inner}$ [M\usun] & 4\e{-6}  & 1\e{-11}     & \ten{-14}-\ten{-5} \\
	  $\Psi_{\rm all,inner}$   & 1.00  & ...    & 0.01 - 1.00  \\ 
	  \hline
	  $\Psi_{\rm all}$         & 0.35  & 0.25        & 0.15-1.0    \\ 
	  \grg [\%]                & 94    & 99 & 0-99.99    \\
	  $\Psi_{\rm big}$         & 0.5            & 1.0         & 0.01-1.0    \\ 
	  A$_v$ [mag]              & 1.2            & 0.95        & 0.0-2       \\ 
	  \hline\hline
	\end{tabular}
	\caption{Stellar and best-fit parameters for the two different geometries.
	  References - 
	  (1) \citet{1996yCat..21010117K}; 
	  (2) \citet{2000ApJ...545.1034S}
	  \label{tab:model_parameters}}
      \end{table}

      \section{Discussion and conclusions}
      We have studied the radiative transfer effects in the gapped disk of LkCa 15 using MCMax, a self-consistent 2D radiative transfer code. We show that only an optically thin \textit{dust shell} or an optically thick \textit{inner disk} can explain the SED in the near and mid infrared. The outer disk geometry differs significantly depending on the inner disk properties. With an optically thick inner disk, the outer disk must rise out of the shadow cast by the inner disk. With an optically thin dust shell, there is no shadow and the outer disk becomes twice as flat.

      These models are similar to but not the same as those proposed in Esp07. The latter do not take into account the shadow cast by an optically thick inner disk on the outer disk. They find the same z/r for the inner disk rim (0.01 AU / 0.12 AU =0.08) as for the outer disk rim (4 AU / 46 AU =0.08). This means that the outer disk wall cannot be illuminated, as it lies in the shadow of the inner disk (see Fig. \ref{fig:LkCa15_SED}, dotted line). The model with an optically thin inner disk can be reproduced, with the exception that to become optically thin, its scale height needs to be so high that it effectively becomes a dust shell.

      Esp08 find the near infrared excess to be a blackbody and conclude that the inner disk has to be optically thick. On the other hand, \cite{2006ApJ...636..348V} concluded that compact dust halos can explain the near infrared bumps in Herbig Ae/Be stars and that their existence agrees well with interferometric data.

      An open question remains as to why the inner disk would have so high a scale height that it becomes an optically thin dust shell. Such a high scale height for small dust grains could be due to a halo of unknown origin, maybe residual infall, but such an infall would not lead to a dust cloud confined to $<$5AU.  Another possibility could be violent scattering and disruption of planetesimals that could create a cloud of dust with a scale height determined by dynamic processes rather than the hydrostatic equilibrium of gas. However, there is currently no additional evidence supporting this scenario for \object{LkCa 15}. Furthermore an optically thick inner disk leads to a model with outer disk settling requirements similar to what is seen in other T Tauri stars.  Therefore we select the model with the optically thick inner disk as the preferred one.

      \begin{figure}
	\centering
	  \includegraphics[width=\linewidth]{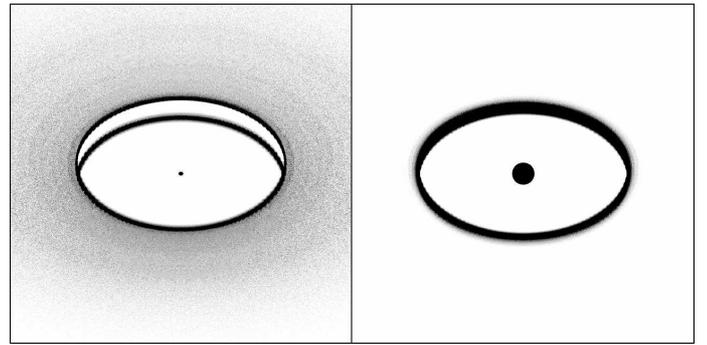} 
	\caption[]{20 micron images of our best-fit disk models of LkCa 15. The left panel shows a model with an optically thick inner disk of 1 AU that casts a shadow on the rim of the outer disk. The right panel shows a 5 AU optically thin dust shell and a fully illuminated outer rim. Note that if the rim of the outer disk is shadowed, much more radiation emerges from the disk surface.\label{fig:LkCa15_shadow}}
      \end{figure}

      To distinguish between geometries with a dust disk or a dust shell - or anything inbetween - a near-infrared interferometer can be used to directly probe the inner regions. But because a shadow from the inner disk connects the inner and outer disk geometry, imaging of the outer disk can also reveal the nature of the inner disk (Fig. \ref{fig:LkCa15_shadow}). If the inner disk is optically thick, radiation from the outer disk will be more extended because the contribution from the shadowed inner rim of the outer disk is small, while the disk surface has a larger flaring angle and contributes to the 20 $\mu m$ flux as well. For an optically thin inner dust shell, the emission will be more centrally peaked because the outer disk is flatter and its rim fully illuminated.

      We can draw one general conclusion about transitional objects: the radial optical depth - and therefore the vertical structure - of dust within an inner hole determines the illumination of the outer disk and ergo its geometry. This is especially important in the light of observations with the Spitzer space telescope, where large numbers of disks with dust-filled inner holes have been identified \citep{2007MNRAS.378..369N}.

      \begin{acknowledgements}
	This research project is financially supported by a joint grant from the
	Netherlands Research School for Astronomy (NOVA) and the Netherlands
	Institute for Space Research (SRON).
	M. Min acknowledges financial support from the Netherlands Organisation for Scientific Research (NWO) through a Veni grant.
      \end{acknowledgements}
      %
      %
      \bibliographystyle{aa} 
      \bibliography{12743} 
      %
      %
\end{document}